\documentclass[useAMS,usenatbib]{mn2e}
\usepackage{epsfig,amssymb,amsmath,graphicx}

\title[Quasi-Periodic Oscillation in the IC 342 X-1]{Discovery of {\bf a} Quasi-Periodic Oscillation in 
the Ultraluminous X-ray Source IC 342 X-1: {\it XMM-Newton Results}}
\author[V. K. Agrawal \& Anuj Nandi]{V. K. Agrawal$^{1}$\thanks{E-mail:
vivekag@isac.gov.in} and  Anuj Nandi$^1$ \\
$^{1}$Space Astronomy Group, SSIF/ISITE Campus, ISRO Satellite Centre, Outer Ring Road, 
Marathahalli, Bangalore 560037, India \\
}
\begin{document}


\pagerange{\pageref{firstpage}--\pageref{lastpage}} \pubyear{2002}

\maketitle

\label{firstpage}

\begin{abstract}

We report the discovery of a quasi-periodic oscillation (QPO) at 642 mHz in an {\it XMM-Newton} 
observation of the ultraluminous X-ray source (ULX) IC 342 X-1. The QPO has a centroid at 
$\nu_{QPO} = 642 \pm 20$ mHz, a coherence factor of $ Q = 11.6$, and an amplitude (rms) of 4.1\% with 
significance of $3.6\sigma$. The energy dependence study shows that the QPO is stronger in the energy 
range 0.3 - 5.0 keV. A subsequent observation (6 days later) does not show any signature of the QPO 
in the power density spectrum. The broadband energy spectra (0.3 - 40.0 keV) obtained by 
quasi-simultaneous observations of {\it XMM-Newton} and {\it NuSTAR} can be well described by an 
absorbed {\it diskbb} plus {\it cutoffpl} model. The best fitted spectral parameters are power-law 
index ($\Gamma$) $\sim$ 1.1, cutoff energy ($E_c$) $\sim$ 7.9 keV and disc temperature 
($kT_{in}$) $\sim$ 0.33 keV, where the QPO is detected. The unabsorbed bolometric luminosity 
is $\sim$ 5.34$\times$ 10$^{39}$ erg~s$^{-1}$. Comparing with the well known X-ray binary GRS 1915+105, 
our results are consistent with the mass of the compact object in IC 342 X-1 being in the range  
$\sim 20 - 65 ~ M_\odot$. We discuss the possible implications of our results.

\end{abstract}


\section{Introduction}

Ultraluminous X-ray sources (ULXs) are off-nuclear point X-ray sources in nearby galaxies with
isotropic luminosities $>$ $10^{39}$ erg~s$^{-1}$ (see \citealt{Feng11} for a recent review). 
Since their discovery more than 30 years ago \citep{Long83, Fab87, Fab88, 
Fab89}, the true nature of ULXs has remained a mystery. Early {\it ASCA} and 
{\it XMM-Newton} observations revealed that most ULXs should contain accreting black 
holes \citep{Kubota01, Sutton13}.
Recently, \citealt{motch14} reported a firm upper limit of $<15$ $M_\odot$ on the mass of a 
black hole in a ULX suggesting that most ULXs are indeed stellar-mass black holes. However, discovery 
of X-ray pulsations \citep{bachetti14} in one of the ULXs in M82 galaxy suggests that ULXs may 
also be powered by accretion onto magnetized neutron stars.
In ULXs, mass estimates of the compact objects remain highly debatable because no dynamical 
measurement has been possible yet. However, recent optical/UV observations
reveal that ULXs in M101 \citep{Liu13} and NGC 7793 \citep{motch14} harbour stellar-mass black holes.

Since luminosities of ULXs exceed the Eddington rate for a 10 $M_\odot$ black hole, it has been
suggested that ULXs might harbour intermediate mass black holes (IMBHs) with mass in the
range $10^2 - 10^4$ $M_\odot$ (\citealt{CM99}; see also \citealt{Pasham14}). Other popular 
models proposed to explain the ultraluminous nature of ULXs are: 
(1) normal X-ray binaries (XRBs) accreting at super-Eddington rate \citep{Begel02}, (2) XRBs 
accreting at sub-Eddington rate with beamed emission \citep{Rey97, King02, Begel06}. However, 
the beaming scenario suffers from several difficulties, e.g, a lack of evidence of 
radio jets in ULXs (see \citealt{Feng11}) and the presence of a strong QPO in M82 X-1 \citep{stroh03}. 
Recently, \citet{Glad09} suggested a new accretion state named the {\it ultraluminous state}  
implying a super-Eddington accretion rate for ULXs. Based on the early observations with 
{\it XMM-Newton}, some ULX spectra (0.3 - 10.0 keV) were modelled with a simple phenomenological 
model (i.e., disc emission and power-law component; \citealt{Miller03, Miller04}) with a characteristic
disc temperature of $kT_{in}$ $\sim$ 0.1 - 0.5 keV. The presence of a cool disk component and
very high luminosities, which indeed triggered the hypothesis that ULXs may contain a IMBH, 
may not be a valid prescription to explain most of the X-ray observational features. Recent 
studies (see, \citealt{Glad09} for details) showed that the energy spectra can be
well described by an optically thick corona ($\tau \sim 5 - 30$, whereas $\tau \sim 1$ for 
corona seen in Galactic black hole binaries) coupled with an accretion disc. In addition, the
presence of a high-energy curvature ($> 3.0$ keV), not seen in Galactic black hole 
binaries \citep{RM06} accreting at a sub-Eddington rate, is now considered as one of the ULX 
spectral signatures \citep{Sto06, Glad09}. Furthermore, combining both spectral and temporal 
variabilities, \cite{Sutton13} classified three spectral regimes for ULXs, namely {\it broadened disc},
{\it hard ultraluminous} and {\it soft ultraluminous} class. Recently, \cite{Pintore14}
investigated a larger sample of observations for studying the spectral evolution with a
different approach (i.e., based on {\it colour-colour} and {\it hardness-intensity diagram} 
analysis). All these studies further corroborate the idea of a new accretion state 
(i.e., {\it ultraluminous state}) for ULXs \citep{Roberts07, Glad09}.


The study of short-term variability may also put constraints on the various accretion models in ULXs 
(see \citealt{Pasham14}).
In a few ULXs (M82 X-1, NGC 5408 X-1, NGC 6946 X-1, M82 X42.3+59), QPOs have been detected in the
frequency range of $\sim$ 3 mHz to 200 mHz \citep{stroh03,Dewa06a,stroh07,Rao10,Feng10,Pasham12}. 
A QPO around 200 mHz has been reported in Holmberg X-1 \citep{Dewa06b}, but was not 
confirmed later (see \citealt{Heil09}). Recently, \cite{Pasham14} reported twin-peak QPOs in 
M82 X-1 at frequencies of 3.32 and 5.07 Hz. Several ULXs show the short-term variability with 
(or without) the presence of QPOs in the power spectra, whereas the variability is found to be 
completely suppressed in some sources \citep{Heil09}. 
It has been suggested that the short-term variability in the ultraluminous state
can be produced by variable obscuration due to clumpy winds (see \citealt{Mid11}; 
\citealt{Sutton13}).


IC 342 X-1 is a ULX in the nearby spiral galaxy IC 342 at a distance of 3.3 Mpc \citep{saha02}. The 
source was discovered by the {\it Einstein} satellite \citep{Fab87}. This source was also detected by
{\it ROSAT} in the ultra-luminous state \citep{Breg93, Roberts00}. {\it ASCA} observations of the
source taken during 1993 and 2000 revealed spectral transitions from a high/soft to a 
low/hard spectrum \citep{Kubota01}. The analysis carried out using {\it Suzaku}, 
{\it XMM-Newton}, {\it Chandra} and {\it Swift} revealed two different power-law (PL) states in 
this source \citep{yoshida13}: low-luminosity PL state and high luminosity PL state.
Recently, \cite{mar14} reported a clear change in a recent {\it Chandra} spectrum with a much
softer spectrum than seen in all previous observations and it has been modelled with a standard 
accretion disc emission.
The source has also been detected in radio using VLA, but the presence of a compact radio jet
was not confirmed in the high spatial resolution data as observed with VLBI
\citep{cseh12}. 

In the present work, we focus on the recent quasi-simultaneous observations of IC 342 X-1 
made by {\it XMM-Newton} and {\it NuSTAR} in August 2012. 
Recently, \cite{Rana14} also analysed the same data sets in order to understand the broadband 
spectral nature of the source. From our detailed analysis (see \S 2 \& \S 3), we report the detection 
of QPO in IC 342 X-1 along with the spectral properties of the source. 

\begin{figure}
\includegraphics[scale=0.28,angle=-90]{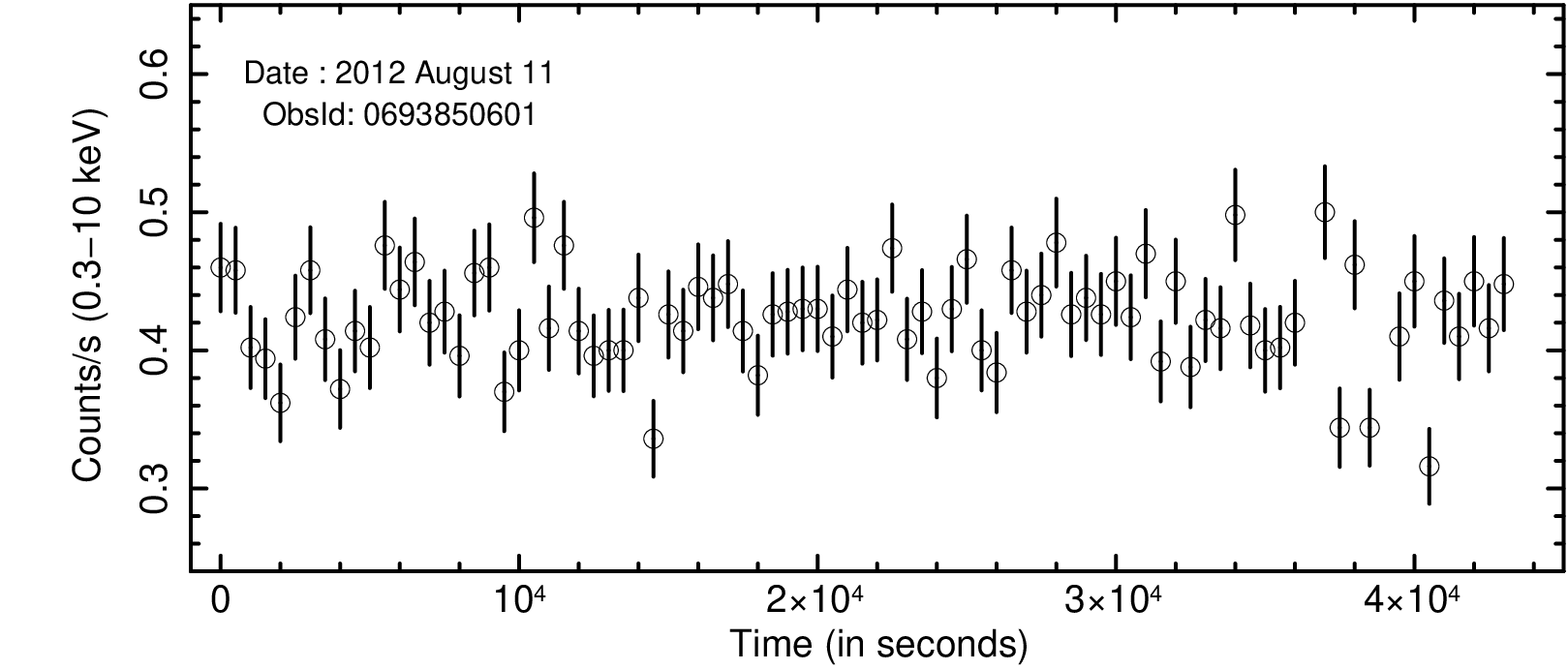}
\includegraphics[scale=0.28,angle=-90]{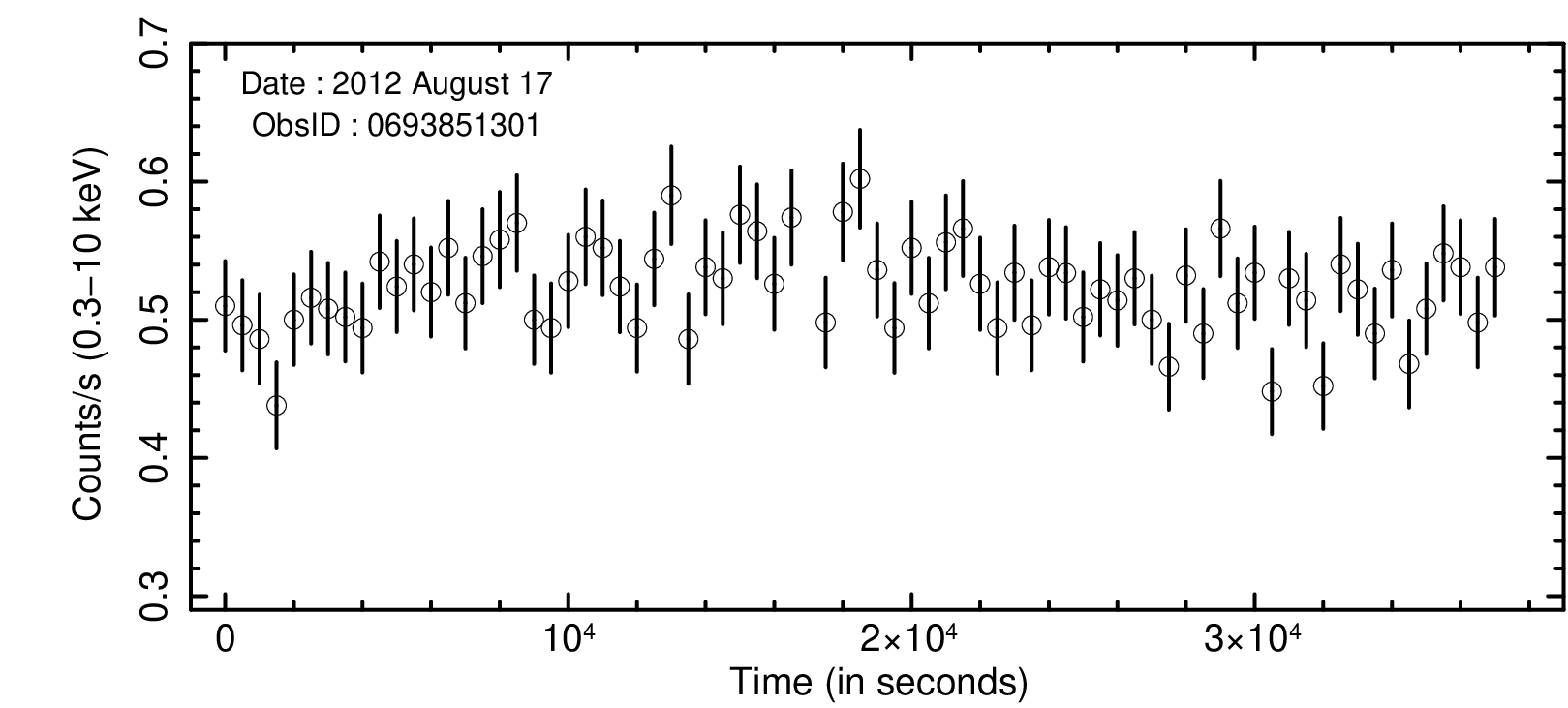}
\caption{Photon counts variation of IC 342 X-1 observed with {\it XMM-Newton} (EPIC-pn data) in the 
energy band of 0.3 - 10.0 keV. {\bf Top panel}: The start time of observation was (Obs-1) 2012-08-11
20h 30m 47s (UT); {\bf Bottom panel}: The start time of observation was (Obs-2) 
2012-08-17 20h 12m 45s (UT). Each data point corresponds to 500 sec time bin.}
\end{figure}

\section{Observation and Data Reduction}

{\it XMM-Newton} observed IC 342 X-1 six (6) times between 2001 and 2012. We use the data 
obtained on 2012 August 11 with an exposure time of 55 ks (Obs-1) and on 2012 August 17 
for a total exposure of 50 ks (Obs-2). The previous data sets (i.e., initial four observations) have 
been analyzed by \cite{yoshida13} (see \citealt{Pintore14} for recent analysis). The 2012 observations 
were carried out in {\it PrimeFullWindow} mode with a time resolution of 73.35 ms. Data 
reduction is performed using Science Analysis System (SAS) Version 12.0.1 and using the recent 
calibration data set. 
We create the calibrated event files using SAS task {\it epchain}. We extract the lightcurve 
from the entire chip in the 10.0 - 15.0 keV band and then generate the gti file using the
selection criteria (rate $<=$ 3 $\times$ the mean of the 10.0 - 15.0 keV lightcurve).
We select EPIC-pn events with $PATTERN<=4$ and $FLAG==0$. We use a circular region of 
$40{\arcsec}$ centered at the source position to extract the source events. A circular region 
of similar size away from the source position is used to extract the background events. 
We apply the gti filter while creating the lightcurves in the 0.3 - 10.0 keV, 0.3 - 5.0 keV and 
5.0 - 10.0 keV bands for Obs-1 and Obs-2. We find that the above 
filtering process removes the large background flares at the end of the observations. We also observe 
that the filtering process produces two data gaps of 500~s and 300~s in the lightcurves of Obs-1 
corresponding to two short flares and {\bf a} 300~s data gap in the lightcurves of Obs-2. 

The task {\it rmfgen} and {\it arfgen} are used to create response matrix file (rmf) and ancillary 
response file (arf). The spectra are grouped to give a minimum of 25 counts/bin. 

{\it NuSTAR} also observed the source on 2012 August 10 for a total exposure time of 98.6 ks and 
2012 August 16 for a total exposure of 127.3 ks. We use the most recent {\it NuSTAR} analysis 
software distributed with HEASOFT version 6.15 and the latest calibration file{\bf s} 
(version 20131007) for reduction and analysis of the {\it NuSTAR} data. We use {\bf the} task 
{\it nupipeline} to generate calibrated and screened event files. A circular region of $30{\arcsec}$ 
centered at the source position is used to extract the source events. Background events are extracted 
from a circular region of same size away from the source. The task {\it nuproduct} is used to 
generate the spectra and response files. The spectra are grouped to give a minimum of 
25 counts/bin.

\section{Analysis and Results}

The background subtracted lightcurves of IC 342 X-1 observed with {\it XMM-Newton} at two different 
epochs (Obs-1 \& Obs-2) are shown in Figure 1 with effective exposure times (resulted after removing 
the high background intervals)  of 43.5 ksec and 37.5 ksec respectively. The average count rate for 
Obs-1 is $0.40 \pm 0.02$ and that for Obs-2 is $0.51 \pm 0.02$. The variability of about 20\% is 
evident in both observations on time scale of few 1000 seconds. The average count rates between 
two observations also vary by $\sim$ 30\%.  

\subsection{Power Density Spectrum}

\begin{figure}
\hskip -.5in
\includegraphics[height=3.9in, width=7.5in, angle=-90]{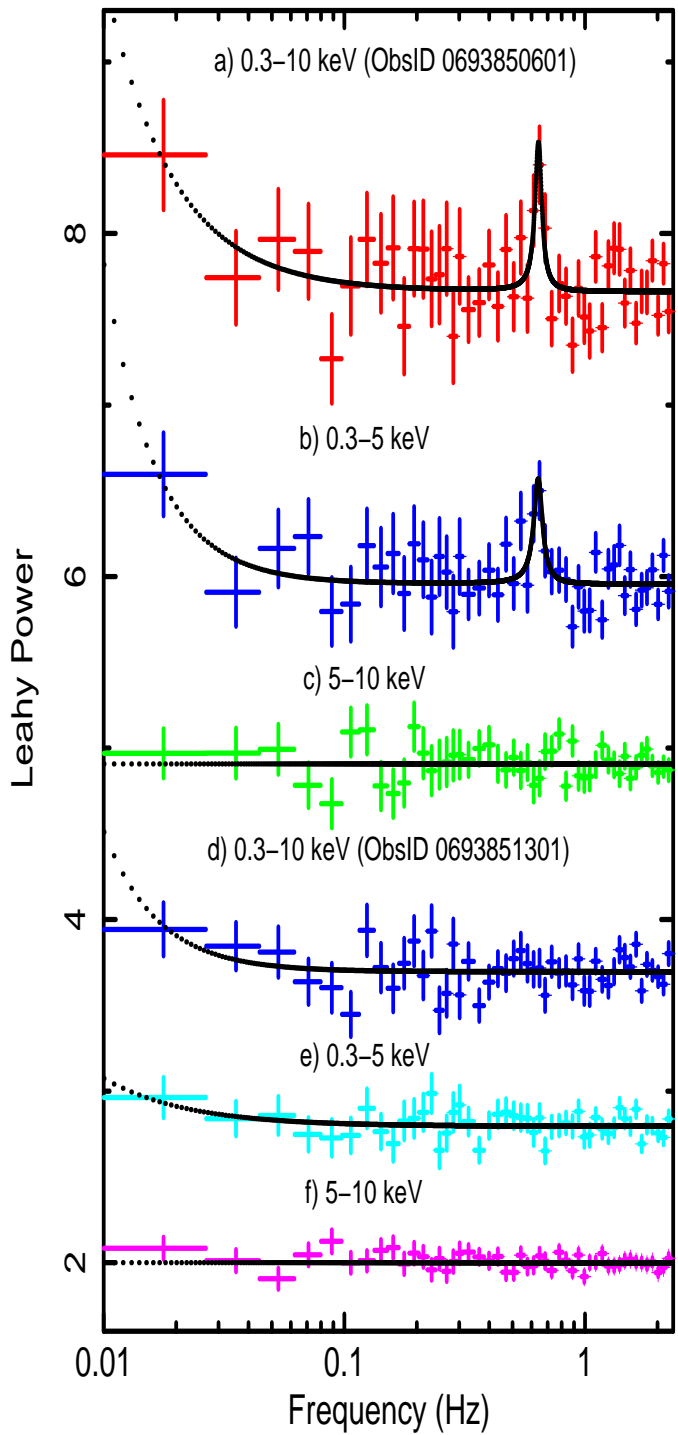}
\caption{{\bf a)} The Leahy normalized power density spectrum for Obs-1 in the 0.3 - 10.0 keV band 
computed using EPIC-pn data. The PDS has been fitted with a Lorentzian for QPO feature, a power-law
and a constant to account for the Poissonian noise. The Poissonian noise has not been subtracted. 
{\bf b)} The Leahy normalized power density spectrum for Obs-1 in the 0.3 - 5.0 keV band. 
A clear QPO feature fitted with Lorentzian is seen. 
{\bf c)} The Leahy normalized power density spectrum for Obs-1 in the 5.0 - 10.0 keV band. 
{\bf d)} The Leahy normalized power density spectrum for Obs-2 in the 0.3 - 10.0 keV band 
without any signature of QPO like feature. 
 {\bf e)} The Leahy normalized power density spectrum for Obs-2 in the 0.3 - 5.0 keV band. 
{\bf f)} The power spectrum for Obs-2 in the 5.0 - 10.0 keV band. See text for details.}
\end{figure}

We compute the power density spectra (PDS) from the background subtracted lightcurves using the 
EPIC-pn data. We follow the procedure given in \cite{Heil09} to construct the PDS. We fill the 
telemetry gaps ($>$ 15 s) and gaps due to the short flares with local averages. We use a
binsize of 0.220 s (3 times the temporal resolution) to construct the PDS. We divide the entire 
lightcurve into intervals of 256 bins (56.32~s) and compute the PDS for each interval independently. 
Then we co-add all the PDS and average them in a single frame. The final PDS is rebinned geometrically
in frequency space by a factor of 1.04.

Figure 2a and Figure 2d show the Leahy power spectra (0.3 - 10.0 keV) for both observations 
(Obs-1 \& Obs-2). The PDS of Obs-1 (Fig. 2a) show{\bf s} a QPO at centroid frequency $\sim$ 642 mHz. 
The PDS of Obs-2 (Fig. 2d) does not show any signature of QPO like feature. 
The PDS in the 0.3 - 5.0 keV range shows QPO at 653 mHz (Fig. 2b). However, no QPO is observed 
in the 5.0 - 10.0 keV power spectrum (Fig. 2c).  

We fit the PDS in the 0.3 - 10.0 keV and 0.3 - 5.0 keV bands (for Obs-1) with a model 
composed of a Lorentzian for a QPO peak, a power-law (AE$^{-\alpha}$, $\alpha$ = 1.5 $\pm$ 0.8 
for the 0.3 - 10.0 keV band, $\alpha$ = 1.95 $\pm$ 0.9 for the 0.3 - 5.0 keV band) and a constant 
to account for the Poissonian noise. We get $\chi_{red}^2$ = 1.02 ($\chi^2 / dof = 45/44$) for the 
0.3 - 10.0 keV band and $\chi_{red}^2$ = 1.06 ($\chi^2 / dof = 47/44$) for the 0.3 - 5.0 keV band. 
The resultant fits are shown in Fig. 2a and Fig. 2b respectively.
Fitting the power spectra (Obs-1) in the 0.3 - 10 keV and 0.3 - 5.0 keV bands  with a power-law
and a constant gives $\chi_{red}^2$ = 1.51 ($\chi^2 / dof = 71/47$) and  $\chi_{red}^2$ = 
1.57 ($\chi^2 / dof = 74/47$) respectively. So, overall fitting is improved upon considering the 
Lorentzian component for the QPO feature. Fitting the power spectra of Obs-2 in the 0.3 - 10.0 keV 
and 0.3 - 5.0 keV bands with a power-law and a constant gives $\chi_{red}^2$ = 1.22 ($\chi^2 / dof 
= 57/47$) and $\chi_{red}^2$ = 1.02 ($\chi^2 / dof = 48/47$) respectively.
We note that no Lorentzian component is required to improve the fit of the PDS of Obs-2 in both 
energy bands (see Fig. 2d, 2e). 
We also note that the PDS in the 5.0 - 10.0 keV energy bands for both observations (Obs-1 and Obs-2) 
are featureless and can be fitted with a constant to account for the Poissonian noise (see Fig. 2c, 2f).
The significance of the QPO parameters in both bands are estimated with F-test statistics. 
The best fit model in the 0.3 - 10.0 keV band gives a QPO of centroid frequency 
$\nu_{QPO} = 642 \pm 20$ mHz, a Q-factor ($\nu/FWHM$) = 11.6 and an amplitude (rms) of 4.1\% with 
significance of 3.6$\sigma$. The best fit QPO parameters in the 0.3 - 5.0 keV band are centroid 
frequency $\nu_{QPO} = 653 \pm 30$ mHz, Q-factor = 9.8 and amplitude (rms) of 4.5\%. The QPO in 
the 0.3 - 5.0 keV band is detected with a significance of 3.7$\sigma$. The total integrated power 
(0.01 - 2.5 Hz) is $\sim$ 7.2\% in the energy band of 0.3 - 10.0 keV for Obs-1 and 3.2\% for Obs-2. 
All errors quoted are computed using $\Delta\chi^2 = 1.0$.

Note that considering the time interval (Obs-1) before the short flares results in 36.5 ks 
continuous observation. The PDS created using this exposure time (36.5 ks) shows a QPO feature with 
similar parameters. However, the significance of the QPO changes from 3.6 to 3.4$\sigma$.

\subsection{Energy Spectrum}

The energy spectra of EPIC-pn and {\it NuSTAR/FPMA} are analysed using XSPEC version 12.8.1. We fit 
the combined EPIC-pn (0.3 - 10.0 keV band; Obs-1) and {\it NuSTAR}/FPMA (3.0 - 40.0 keV; 2012 
August 10) data (epoch-1) with: 1)  {\it power-law}, 2) power-law with exponential cutoff
({\it cutoffpl} in XSPEC), 3) {\it cutoffpl} with an addition of multi-temperature disk 
component for the standard thin accretion disk ({\it diskbb} in XSPEC; \citealt{Mit84}), 
4) {\it diskbb} plus Comptonization model ({\it compTT} model of XSPEC; \citealt{Titar94}). We 
consider the {\it tbabs} model \citep{Wilms00} for all the spectral models in order to model the 
extinction ($N_H$) on the line of sight to the source. Similarly, the second quasi-simultaneous data 
(epoch-2) obtained with EPIC-pn (0.3 - 10.0 keV; Obs-2) and {\it NuSTAR}/FPMA (3.0 - 40.0 keV; 2012 
August 16) are also analyzed and fitted with above mentioned models. The cross-calibration constant 
between {\it NuSTAR/FPMA} and {\it XMM-Newton/EPIC-pn} is found to be close to 1 ($\sim$ 0.95).

The best fit parameters for all the models are listed in Table 1. All errors quoted are computed 
using $\Delta\chi^2 = 1.0$ (at 68\% confidence). The $N_H$ values are found to be in the range of 
0.5 - 0.7 $\times$ 10$^{22}$ $cm^{-2}$(see Table 1). For epoch-1, the {\it power-law} and 
{\it cutoffpl} model give $\chi_{red}^2 = 1.22$ ($\chi^2/dof$ = 785/643) and $\chi_{red}^2 = 1.10$ 
($\chi^2/dof$ = 706/642) respectively. The probability that the fit is improved by chance 
is 1.59 $\times$ 10$^{-16}$. Hence a simple {\it power-law} model is not the best description of 
the data and the spectrum shows a clear high energy cutoff. Since {\bf a} soft excess modelled 
with the {\it diskbb} has been observed in many ULXs \citep{Miller03,Miller04, Feng05,  Sto06}, we 
fitted the epoch-1 spectrum with the {\it diskbb+cutoffpl} model. 
The fit results in $\chi_{red}^2 = 0.99$ ($\chi^2/dof$ = 634/640) and chance improvement probability 
equals to 1.12$\times$ 10$^{-15}$ for inclusion of the {\it diskbb} component, 
suggesting that a soft excess component below 2 keV is required to improve the fit.

Analysis of XMM-Newton data of several ULXs revealed that the disc emission plus cool 
($kTe \sim $ 3 keV) and optically thick ($\tau \sim$ 5 - 30) Comptonized component models the 
spectral data well \citep{Glad09}. Hence, we tried with the {\it compTT} model also instead of 
{\it cutoffpl}. The combination of {\it diskbb} plus {\it compTT} model, where the seed photon
temperature is tied to the inner disc temperature, gives $\chi_{red}^2 = 1.01$ ($\chi^2/dof$ =
645/640). Hence, the {\it diskbb+cutoffpl} model is the best description for the epoch-1 observation. 
epoch-2 data can also be well modelled with the {\it diskbb+cutoffpl} model resulting in $\chi_{red}^2 
= 1.04$ ($\chi^2/dof$ = 729/699).

\begin{figure}
\includegraphics[height=8.2cm,width=5.0cm,angle=-90]{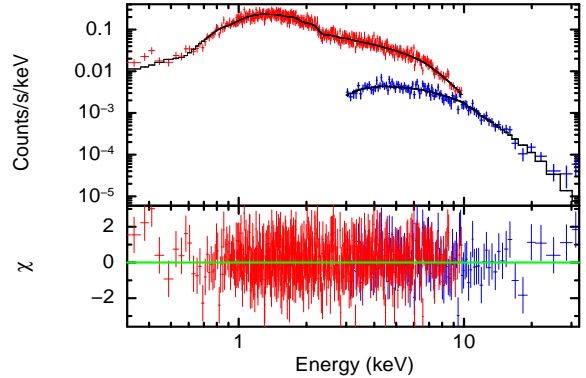}
\caption{The combined EPIC-pn (for Obs-1) and {\it NuSTAR/FPMA} (on August 10 2012) spectrum and 
folded model (top panel). The spectrum has been modeled with an absorbed {\it diskbb} plus 
{\it cutoffpl} model. The residuals in unit of sigma is shown in the bottom panel.}
\end{figure}

The disc temperature is 0.33 $\pm$ 0.03 keV and 0.47 $\pm$ 0.06 keV respectively for the epoch-1 and 
epoch-2 data. The {\it diskbb} normalization is found to be unphysically small for both
spectra. The cutoff energy ($E_c$) is 7.92 $\pm$ 0.91 keV and 7.27 $\pm$ 0.99 keV respectively 
for epoch-1 and epoch-2 spectra (see also \citealt{Rana14}). Although the {\it diskbb+cutoffpl} 
model is the best description of the spectra, the {\it diskbb+compTT} model also provides 
statistically good description of the spectra for both epochs. The best fit optical depth ($\tau$) 
and electron temperature ($kT_e$) for the epoch-1 spectrum are 13.32 $\pm$ 0.66 and 3.35 $\pm$ 0.17 keV 
respectively. The best fit $\tau$ and $kT_e$ for the epoch-2 are 12.31 $\pm$ 0.41 and 
3.29 $\pm$ 0.15 keV respectively. 

In Table 1, we also give the estimated unabsorbed total flux in 0.1 - 100.0 keV band and 
corresponding luminosities for all the models using a distance of 3.3 Mpc. The unabsorbed 
fluxes and corresponding errors are computed using the convolution model {\it cflux} of XSPEC.

\section{Discussion And Summary}

In the present work, we report the discovery of a QPO in the {\it XMM-Newton} data of the ULX 
source IC 342 X-1, a system harbouring a black hole (see \citealt{Okada98}).
The detection of a peak at $\sim$ 642 mHz (Q $\sim$ 11.6, rms $\sim$ 4.1\% and significance 
$3.6\sigma$) in the power spectrum of IC 342 X-1 (Obs-1) could be used to constrain the mass of 
the `hole' as this technique was employed for other ULXs \citep{Dewa06a, stroh07, Rao10}. 
Interestingly, the QPO detection in IC 342 X-1 is also the highest frequency observed in a ULX 
to date, whereas the subsequent {\it XMM-Newton} observation (6 days later) does not 
show any signature of QPO like feature.   

\begin{table*}
\scriptsize
\caption{Summary of the spectral fits ($\Gamma$ is photon index, $E_c$ is cutoff energy, 
$kT_{in}$ is disk temperature, $N_{dbb}$ is disk normalization, $kT_e$ is electron temperature of 
corona, $\tau$ is optical depth of corona. $F_{tot}$ is total flux in units of 10$^{-12}$ erg/s/cm$^2$, 
$L_{tot}$ is total luminosity in units of 10$^{39}$ erg/s). Errors quoted are calculated
at 68\% confidence level.}
\centering
\begin{tabular}{|l|l|l|l|l|l|l|l|l|} \hline
&  \multicolumn{4}{|c|}{epoch-1} & \multicolumn{4}{|c|}{epoch-2} \\ \hline
Parameters & power-law & cutoffpl  &                           diskbb+cutoffpl       &diskbb+compTT&      power-law &         cutoffpl     & diskbb+cutoffpl & diskbb+compTT\\ \hline
$N_H$ ($\times$ 10$^{22}$ cm$^{-2}$)         &0.61$\pm$0.01     &  0.54$\pm$0.01   & 0.64$\pm$0.03 & 0.67$\pm$0.06    & 0.71$\pm$0.03   &0.61$\pm$0.02 & 0.59$\pm$0.02 &0.51$\pm$0.04\\
$\Gamma$     & 1.82$\pm$0.02    & 1.51$\pm$0.04    &1.04$\pm$0.10   & $-$             & 2.04$\pm$0.02    &1.66$\pm$0.04     &1.12$\pm$0.18 & $- $ \\
$E_c$ (in keV)       & $-$              &15.93$\pm$1.7      &7.92$\pm$0.9    & $-$             &  $-$             &13.45$\pm$1.2     & 7.27$\pm$0.95& $ - $ \\
$kT_{in}$ (in keV)    &  $-$             &$-$               &0.33$\pm$0.03   & 0.23$\pm$0.03   &  $-$             &$-$               & 0.47$\pm$0.06& 0.29$\pm$0.02 \\
$N_{dbb}$     &   $-$             &$-$               &4.96$^{+3.8}_{-1.1}$ & 28$^{+30.8}_{-10.4}$ &  $-$    &$ - $           & 1.08$\pm$0.3& 3.12$^{+3.01}_{-1.21}$\\
$kT_e$  (in keV)     &   $-$            &$-$               &$-$             & 3.35$\pm$0.17            &$-$       &$-$   &         $-$             & 3.29$\pm$0.15 \\
$\tau$       &   $-$            &$-$               &$-$             & 13.32$\pm$0.66          &$-$        & $-$            &  $-$   &       12.31$\pm$0.41  \\
$F_{tot}$ (0.1-100 keV)  &8.91$\pm$0.08            &5.05$\pm$0.06              &4.21$\pm$0.11            &4.46$\pm$0.05       & 8.49$\pm$0.12 &5.88$\pm$0.09            &4.21$\pm$0.2 &     4.36$\pm$0.3         \\
$L_{tot}$ (0.1-100 keV) &11.56$\pm$0.01             &6.55$\pm$0.07              &5.34$\pm$0.14            & 5.78$\pm$0.06   &    11.01$\pm$0.02            &7.62$\pm$0.01 &    5.31$\pm$0.25  &5.65$\pm$0.26     \\
$\chi^2$/dof    & 785/643      & 706/642 &         634/640 &         645/640                 & 884/702 & 756/701           & 729/699 & 736/699\\
\hline
\end{tabular}
\end{table*}



Detailed spectral analysis shows that the broadband spectrum (0.3 - 40.0 keV) of the source 
(epoch-1) is well described by an absorbed {\it diskbb} plus a {\it cutoffpl} or with a {\it diskbb} 
plus a {\it compTT} model. The best fit model parameters indicate that the source was in a hard spectral 
state ($\Gamma \sim 1.04$, $\tau$ $\sim$ 13.32, $E_c$ $\sim$ 7.92 keV) that is unlike the 
canonical hard state as commonly seen in Galactic black hole binaries (GBHBs). The unabsorbed 
luminosity of the source is found to be around $\sim 5.34 \times 10^{39}$ erg~s$^{-1}$.
Similar spectral features are also seen in the epoch-2 observation. All these observational results 
are consistent with the findings of \cite{Glad09} and implies that the source could be in the hard 
ultraluminous state \citep{Sutton13}. 


In general, the power spectra of GBH sources are characterized by various broad noise 
components (power-law like red noise, flat-top noise etc.) along-with a narrow noise component 
(i.e. Lorentzian type for QPO like feature). The high temporal variability and low frequency QPOs 
($\sim 0.1 - 20$ Hz) observed in BH sources are mostly associated with the hard or 
intermediate states. In general, the low frequency QPOs seen in Galactic BHs are classified 
as A, B and C type \citep{CBS05}. As an example, the C-type QPO has a high Q-factor (6 - 12), a 
large amplitude (3\% - 16\% rms) and shows complex phase lag behaviour along with flat-top noise 
component. Moreover, the low frequency QPOs evolve with the spectral states and completely 
disappear in the thermally dominated soft state (\citealt{RM06}; \citealt{Nandi12}).
Since the QPOs scale inversely with the mass of a black hole in GBH sources \citep{sh03,st09}, the 
observed QPOs can be used to infer the mass of the putative black hole. 

In contrast, the nature of the power spectra and short-term variability are more complex in ULXs 
\citep{Heil09}. Few sources show signature{\bf s} of QPOs without any change in the centroid 
QPO frequency (except M82 X-1, see \citealt{Feng11} for details). The recent discovery of twin-peak
X-ray QPOs (3:2 frequency ratio) in M82 X-1 is found to be stable (see \citealt{Pasham14}). 
No definite correlation exists between the QPOs and spectral states in ULXs, but a defined 
correlation exists for most GBHBs. 
In the case of IC 342 X-1, the power spectrum of Obs-1 can be modelled with a constant and a 
Lorentzian feature (for QPO) along with a power-law component. The observed intrinsic 
variability is around $\sim 7.2\%$. This form of PDS is a common characteristic of 
ULXs (see \citealt{Heil09}), where significant variability is observed and the power spectra are 
modelled with a power-law or with a broken power-law component. Interestingly, the power 
spectrum of Obs-2 (6 days later) does not show any presence of a QPO but both observations (Obs-1 
\& Obs-2) show similar spectral nature (see Table 1). 


Another important thing to note {\bf is} that the hard X-ray energy spectra of GBHBs (i.e., hard 
or intermediates states) have connection with the strong QPOs and compact radio jets observed in 
the sources \citep{Rad14}. The high energy spectral curvatures observed in GBHBs (in the range 
of 30 - 100 keV) are the manifestation of optically thin corona with characteristic temperature 
of $\sim 100$ keV 
(\citealt{RM06}). However, in ULXs the spectral nature is quite different with spectral curvature 
of electron temperature of few keV and optical depth in the range of 5 to 30 (thick corona). Also, 
the connection between the hard X-ray spectral nature, the QPOs and compact jets is not well 
established in ULXs. So, it is hard to compare the spectral nature of ULXs with that of GBHBs. 

Since the characteristic time-scale of active galactic nuclei (AGNs) or a GBH scales with the 
compact object mass (McHardy et al. 2006),
one can raise the following basic question - could the observed QPO in IC 342 X-1 be analogous to 
any type of QPOs observed in Galactic black holes? 
There have been several attempts to identify the QPOs seen in ULXs \citep{stroh09,Pasham12,Feng10} 
with those seen in GBHBs. However, the classification of QPOs observed in ULXs remains 
unclear \citep{Mid11}.
The detection of 642 mHz QPO in IC 342 X-1 which shows some of the characteristics (Q-factor = 11.6 
and rms = 4.1\%) of C-type QPO, can be used to constrain the black hole mass in IC 342 X-1. However, 
true classification requires phase lag study which is beyond the scope of the present work. The 
heaviest Galactic BH source GRS 1915+105 \citep{GCM01} shows C-type 
QPOs in the frequency range 1 - 3 Hz. If we consider that the 642 mHz QPO is a scaled-down 
version of the QPOs seen in GRS 1915+10 along with the assumption that the QPO frequencies 
scale inversely proportional to the BH masses \citep{RM06}, then we can estimate the black hole mass 
in IC 342 X-1 as 
$M_{BH} (IC 342 X-1) \sim \nu_{QPO}(GRS 1915)/\nu_{QPO}(IC 342 X-1) \times M_{BH} (GRS 1915) 
\sim 20 - 65 ~ M_\odot$ (considering the mass of GRS 1915+105 $\sim$ 14 $M_{\odot}$).
It implies that for the bolometric luminosity $\sim 5.34 \times 10^{39}$ erg~s$^{-1}$, the 
black hole at the centre of IC 342 X-1 is accreting matter at near-Eddington rate 
($\sim 0.7 - 2~L_{Edd}$). 






Another possibility for estimating the mass of the black hole in IC 342 X-1 is to consider the 
observed QPO as a scaled-down version of those HFQPOs which are observed in GBHBs (eg. 65-67 Hz 
in GRS 1915+105; see \citealt{MRG97}; 66 Hz in IGR J17091-3624; see \citealt{AB12}).
In this scenario, the estimated mass could be in the range of 
$\sim 1000 - 1800~M_\odot$ (considering the mass of GRS 1915+105 $\sim$ 14 $\pm$ 4 $M_{\odot}$). 
It would imply that the central black hole accretes matter at sub-Eddington rate (0.02 - 0.05 $L_{Edd}$).
This may not be the case for IC 342 X-1, as the observed spectral properties favour the 
{\it hard ultraluminous state} of ULXs, explained by a system harbouring a stellar-mass black hole 
accreting at and above the Eddington limit \citep{Glad09, Sutton13}.

We are then left with the possibility that the central `hole' of IC 342 X-1 might be harbouring 
a `massive' stellar-mass black hole of mass $\sim 20 - 65~M_\odot$. The present stellar evolution 
models also predict massive BH remnants ($\sim 20 - 100~M_\odot$) that could be formed from 
direct collapse of the progenitor without any supernova explosion \citep{Fry99}.





\section*{Acknowledgments}
Authors would like to thank the anonymous referee for his/her careful reading of the paper, and 
constructive criticism along with useful suggestions for improvement of the manuscript.
We also would like to thank Dr. Girish and Dr. Das for careful reading and comments on the
manuscript. This research has made use of data and/or software provided by the High Energy Astrophysics 
Science Archive Research Center (HEASARC).
We thank Dr. Anil Agarwal, GD, SAG, Mr. Vasantha E. DD, CDA and Dr. S. K. Shivakumar, Director, ISAC 
for encouragement and continuous support to carry out this research.

\end{document}